\documentclass[preprint,notitlepage,nofootinbib]{revtex4}
\usepackage{ulem}
\usepackage{amssymb, amsmath, amsopn, color, graphicx}
\usepackage[english]{babel}

\usepackage{bm}
\DeclareMathAlphabet{\matheul}{U}{eus}{m}{n}
\usepackage{float}

\begin{document}


\title{An exact time-dependent interior Schwarzschild solution}
\thanks{}%

\author{Philip Beltracchi and Paolo Gondolo}
\email{phipbel@aol.com, paolo.gondolo@utah.edu}
\affiliation{Department of Physics and Astronomy, University of Utah, 115 South 1400 East Suite 201, Salt Lake City, UT 84012-0830}

\collaboration{}

\date{\today}

\begin{abstract}
We present a time-dependent uniform-density interior Schwarzschild solution, an exact solution to the Einstein field equations. Our solution describes the collapse (or the time-reversed expansion) of an object from an infinite radius to an intermediate radius of $9/8$ of the Schwarzschild radius, at which time a curvature singularity appears at the origin, and then continues beyond the singularity to a gravastar with radius equal to the Schwarzschild radius.
\begin{description}
\item[Usage]

\item[PACS numbers]
 
\item[Structure]
\end{description}
\end{abstract}

\pacs{Valid PACS appear here}
\maketitle

\section{Introduction}
 Nearly 100 years after its original discovery, the constant density interior Schwarzschild solution \cite{Schwarzschild:1916ae} was analyzed in more detail and shown to behave as a gravastar in the limit that the radius approaches the Schwarzschild radius $R_S=2GM$ \cite{0264-9381-32-21-215024}.  If the radius reaches $9/8R_S$  the pressure at the center diverges and the convention was this implied a static solution no longer existed \cite{Carroll:2004st,Schwarzschild:1916ae}. However, the static interior Schwarzschild solution may be maintained, without modification, if one accepts a region of negative pressure \cite{Som200150}.
 While the interior Schwarzschild solution strictly speaking does not avoid singularities (it is singular where the pressure diverges~\cite{cattoen2005gravastars}) it is a simple and mathematically valid solution with potential for high compactness and negative pressures that can be interesting to study in its own right. For example, slowly rotating Schwarzschild stars in the compact gravastar limit behave almost exactly as non black hole extended Kerr sources \cite{posada2017slowly}. Also, if one allows for a Dirac delta function in the transverse stress at the radius of the pressure divergence, the singularity has a well-defined contribution to the Komar integral \cite{0264-9381-32-21-215024}.
 
 In this paper, we show that if one allows for a time-dependent radius and for anisotropic stress, the interior Schwarzschild solution generalizes into a new exact solution to the Einstein field equations. Its line element has the same form as the static interior Schwarzschild solution except for a time-dependent radius $R(t)$. For $r<R(t)$,
 \begin{align}
     ds^2=
    -\frac{1}{4} \left( 3 \sqrt{1 - \frac{R_S}{R(t)} } -  \sqrt{1 - \frac{R_S r^2}{R(t)^3} }\right)^2  dt^2+\left(1 - \frac{R_S r^2}{R(t)^3}\right)^{-1} \, dr^2+r^2d\theta^2 + r^2 \sin^2\theta  \, d\phi^2,
\end{align}
and for $ r\ge R(t)$,
\begin{align}
      ds^2= -\left( 1 - \frac{R_S}{r} \right) \, dt^2+\left(1 - \frac{R_S}{r}\right)^{-1} \, dr^2+r^2d\theta^2 + r^2 \sin^2\theta  \, d\phi^2.
 \end{align}
 The time-dependence of the radius is implicitly defined as the solution to the equation
 \begin{equation}
 \frac{1}{7a^7}-\frac{2}{5a^5}+\frac{1}{3a^3} =\alpha t+\beta, \qquad\text{where } a=\sqrt{1-\frac{R_S}{R(t)}}
 \label{easysol}
 \end{equation}
 and $\alpha$ and $\beta$ are determined by the choice of time origin and collapse or expansion time scale.
\section{ Energy- momentum tensor}

The following functions will be used for shortening some expressions:
\begin{align}
 a = \sqrt{1 - \frac{R_S}{R} } ,
 \qquad
 b = \sqrt{1 - \frac{R_S r^2}{R^3} } ,
 \qquad
 R = R(t).
\end{align}
In terms of $a$ and $b$ the interior metric reads
\begin{equation}
    ds^2= -\frac{1}{4} \left( 3a -  b\right)^2  dt^2+b^{-2}dr^2+r^2d\theta^2 + r^2 \sin^2\theta  \, d\phi^2.
\end{equation}
In the interior Schwarzschild solution the radius $R$ is always greater than the Schwarzschild radius $R_S$.
The energy-momentum tensor for the standard time-independent interior Schwarzschild solution contains a constant density and equal pressures in the radial and transverse directions and no off-diagonal terms. For the time-dependent solution, the energy-momentum tensor $T_{\mu\nu}$ demanded by the Einstein equations $G_{\mu\nu} = 8 \pi G T_{\mu\nu}$ is slightly different. One can isolate the energy density and pressures on the diagonal by raising an index, but the off-diagonal terms are no longer symmetric. Alternatively, one can use tetrads to find the energy tensor in a local Lorentz frame. In this way, one fixes the functions on the diagonal to their correct value and also keeps the off-diagonal terms symmetric. Introducing a tetrad
\begin{equation}
    e^\mu_{\hat{\mu}}=
\begin{pmatrix}
 \frac{2}{|3a-b|} & 0 & 0 & 0 \\
 0 & b & 0 & 0 \\
 0 & 0 & \frac{1}{r} & 0 \\
 0 & 0 & 0 & \frac{1}{r\sin\theta} \\
\end{pmatrix},
\end{equation}
such that
\begin{equation}
 e^\mu_{\hat{\mu}}e^\nu_{\hat{\nu}}g_{\mu \nu}=\eta_{\hat{\mu}\hat{\nu}}=\mathop{\rm diag}(-1,1,1,1),\\
\end{equation}
we have
\begin{equation}
e^\mu_{\hat{\mu}}e^\nu_{\hat{\nu}}T_{\mu \nu}=T_{\hat{\mu}\hat{\nu}} = \, \begin{pmatrix}
\rho & -S_r & 0 & 0 \\
-S_r & p_r & 0 & 0 \\
0 & 0 & p_T & 0 \\
0 & 0 & 0 & p_T 
\end{pmatrix}.
\end{equation}
The energy-momentum tensor can be brought to canonical from and it is either type I or type IV depending on the sign of $(\rho+p_r)^2-4 S_r^2$ (positive or 0 for type I, negative for type IV).
The energy density $\rho$ and radial pressure $p_r$ assume the standard expressions found in the literature~\cite{Carroll:2004st,Som200150,0264-9381-32-21-215024}, although with a time-dependent radius, 
\begin{align}
\rho\equiv -T^0_{\,\,\,0}=T_{\hat{0} \hat{0}} = \frac{3 M}{4 \pi  R^3} ,\qquad
  p_r \equiv T^1_{\,\,\,1}=T_{\hat{1} \hat{1}}= \rho \, \frac{b-a}{3a-b}. 
  \label{rhoandpr}
\end{align}
The radial pressure $p_r$ diverges where 
\begin{equation}
    r=R\sqrt{9-8R/R_S}.
\end{equation}
The $T_{\hat{\mu}\hat{\nu}}$ components that differ from the static case are a new radial momentum flux term $S_r$ and the tangential pressure $p_T$,
\begin{align}
S_r=-T_{\hat{1} \hat{0}}=-T_{\hat{0} \hat{1}}=\frac{2 \rho r \dot{R}}{ b |3 a-b| R},\qquad 
p_T \equiv T^2_{\,\,\,2} = T^3_{\,\,\,3} =T_{\hat{2} \hat{2}}=T_{\hat{3} \hat{3}}= p_r + \Delta.
\end{align}
Here $\dot{R}=dR/dt$ and the pressure anisotropy $\Delta$ follows from Einstein's equations as
\begin{align}
\Delta =  \frac{2 \rho r^2 b R^3 }{2 \pi  (3 a-b)} \frac{\partial }{\partial t}\bigg[\frac{\dot{R}}{(3 a-b) b^3
   R^4}\bigg].
\label{eq:ptA}
\end{align}
All other terms in the energy-momentum tensor are 0 as they must be for a spherical symmetric time-dependent system.

The static solution is recovered for $\dot{R}=0$, and it has $S_r=0$ and $\Delta=0$, i.e., isotropic pressure $p_r=p_T$.
At the outer boundary $r=R(t)$, the energy density jumps from $\rho$ inside to zero outside, the radial pressure $p_r(R)$ vanishes and is continuous, and the tangential pressure $p_T$ assumes the expression
\begin{align}
p_T(R) = \Delta(R) = - \rho \, \frac{{\dot{R}}^2 (2 G M+8 R)+2 R \ddot{R} (2 G M-R)}{4  R (1-2 G M/R)^3}.
\end{align}
A natural  boundary condition is for the tangential pressure to be continuous at the surface, i.e., to set $p_T(R)=0$. This leads to the differential equation
\begin{equation}
   2 R \ddot{R} (2GM-R)+{\dot{R}}^2 (2GM+8 R)=0 .
   \label{eq:diffeq}
\end{equation}
We use this equation to determine the time dependence of the radius $R(t)$.
\section{Time dependence of the radius}
\begin{figure}[H]
    \centering
    \includegraphics[width=7cm]{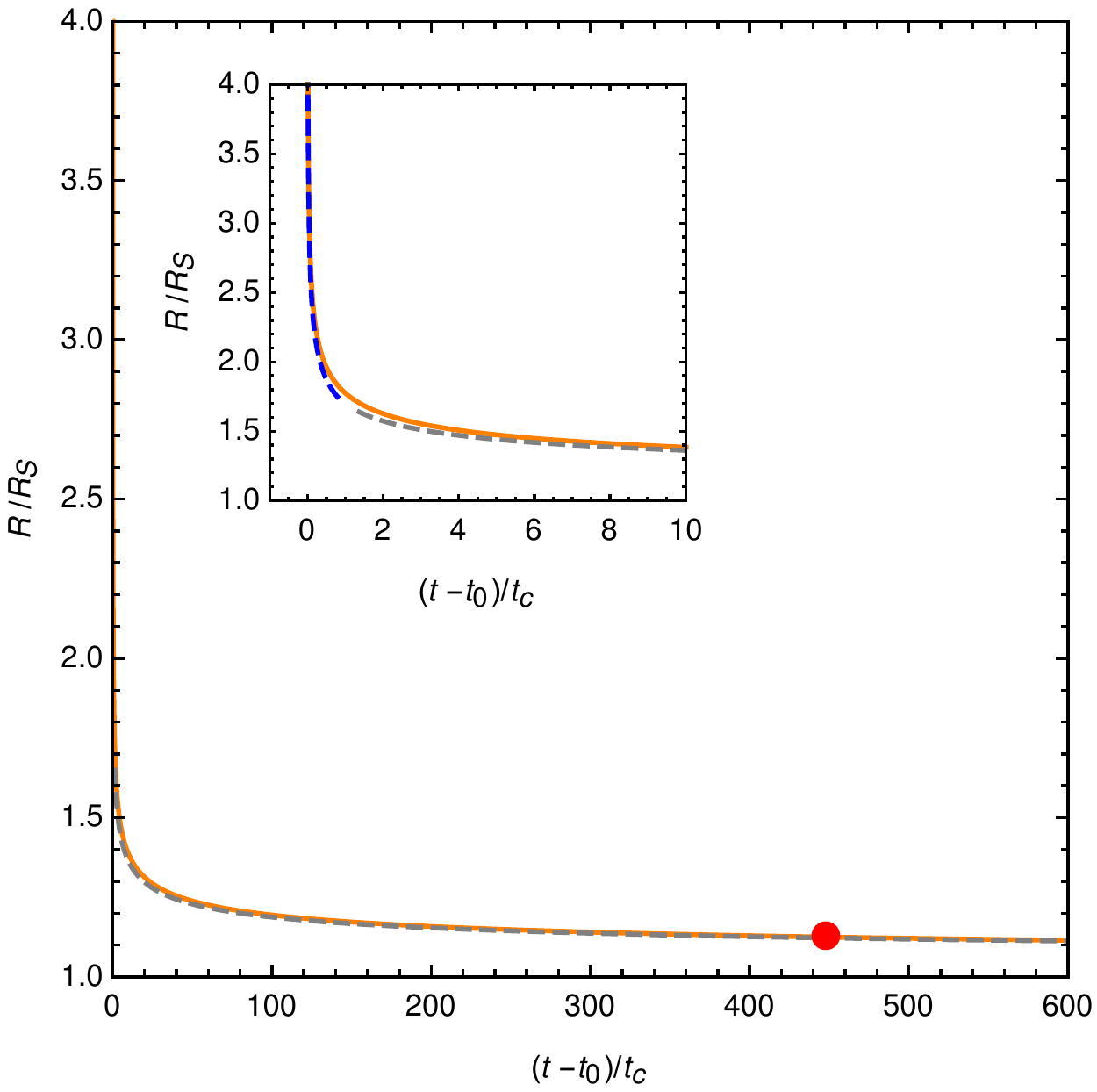}
    \caption{Time dependence of the radius $R(t)$. The solid (orange) line is the full solution. The dashed (blue and gray) lines are the early and late time approximations. The (red) point marks the central pressure divergence at $\tilde{R}=9/8$, $(t-t_0)/t_c=47072/105$. The plot has an inset region showing earlier times in more detail.}
    \label{timeevo}
\end{figure}

The static solution $\dot{R}=\ddot{R}=0$, or constant $R$, satisfies Eq.~(\ref{eq:diffeq}), as expected. We find that an additional time-dependent solution exists.
The general solution to Eq.~(\ref{eq:diffeq}) can be found by reexpressing it as an equation for the function $t(R)$ instead of the function $R(t)$ using the formulas for derivatives of inverse functions $dt/dR=1/\dot{R}$ and $d^2t/dR^2=-\ddot{R}/{\dot{R}^3}$. This leads to the linear differential equation for $t(\tilde{R})$,  where $\tilde{R} = R/R_S$,
\begin{equation}
\label{eq:lde}
\left(8\tilde{R}+1\right) \, \frac{dt}{d\tilde{R}} + 2 \left( \tilde{R}-1 \right) \tilde{R}\, \frac{d^2t}{d\tilde{R}^2} = 0.
\end{equation}
The general solution to Eq.~(\ref{eq:lde}) can be written as 
\begin{align}
   \label{eq:t(R)}
   \frac{t-t_0}{t_c} =\frac{2}{105}\Bigg[\frac{\tilde{R}^{3/2} \left(8 \tilde{R}^2-28 \tilde{R}+35\right)}{(\tilde{R}-1)^{7/2}}-8\Bigg],
   \end{align}
where $t_0$ and $t_c$ are integration constants. This coincides with Eq.~(\ref{easysol}) in the introduction when $\alpha=\frac{1}{2t_c}$ and $\beta=\frac{8}{105}-t_0/(2t_c)$.
The time $t_0$ is the time at which $R=\infty$. The time $t_c$ sets the time scale for the collapse ($t_c>0$) or expansion ($t_c<0$).
While  Eq.~(\ref{eq:t(R)}) is an implicit equation for $\tilde{R}$, it is possible to find asymptotic expressions for $\tilde{R}(t)$ in the regimes $t\rightarrow t_0$ and $t\rightarrow \pm \infty$,
\begin{align}
 \tilde{R}(t)\approx1+\left(3\frac{t-t_0}{t_c}\right)^{-1/3},\quad t\rightarrow t_0;\qquad\tilde{R}(t)\approx1+\left(\frac{7}{2}\frac{t-t_0}{t_c}\right)^{-2/7},\quad t\rightarrow \pm \infty.
 \end{align}
 Plots of the exact and asymptotic solutions for $\tilde{R}(t)$ are shown in Fig.~\ref{timeevo}.
 For reference, our solution gives
\begin{equation}
\frac{d\tilde{R}}{dt}  = -\frac{ (\tilde{R}-1)^{9/2}}{\sqrt{\tilde{R}} t_c},\qquad \frac{d^2\tilde{R}}{dt^2}  = \frac{ (\tilde{R}-1)^8 (8 \tilde{R}+1)}{2 \tilde{R}^2 t_c^2} .
\end{equation}
These expressions lead to
\begin{align}
&\Delta = 
3\rho \, r^2  \tilde{R}^6 a^{16} \, \frac{  (b-a)(9a^2+5ab-3b^2-a^2b^2) }{ (3a-b)^3 \, b^4 \, t_c^2 }\label{eq:del},\\   & S_r=-\frac{2 r \rho a^9 \tilde{R}^3}{|3a-b|bt_c} \label{eq:Sr} .
\end{align}
 The pressure at $r=0$ diverges when $\tilde{R}=9/8$, which happens when $ \frac{t-t_0}{t_c}=47072/105\approx448.3$.
 On the surface of the star ($r=R$), one has $b=a=(1-R_S/R)^{1/2}$ and $\Delta=0$, as imposed.
As $t\to\infty$ for collapse (or $-\infty$ for expansion), the star radius $R\to R_S$, the density becomes $\rho_S =3 M/(4 \pi R_S^3)$ for $r<R_S$ and zero otherwise, the pressure becomes $p_r=p_T=-\rho_S$ for $r<R_S$ and zero otherwise. So our solution describes collapse (expansion) of a constant-density anisotropic object ending (starting) as a sphere with vacuum equation of state $p_r=p_T=-\rho$ and radius equal to the Schwarzschild radius. 

\section{Analysis}
In this section we analyze the metric functions $g_{t t}(t,r)$, $g_{r r}(t,r)$, $\sqrt{-g(t,r)}$, where $g=\det(g_{\mu \nu})$, the Ricci scalar, and the matter functions $\rho(t,r)$, $p_r(t,r)$, $\Delta(t,r)$, and $S_r(t,r)$, paying particular attention to their singularities. Since $R(t)$ is a continuous function of $t$, we study the metric, curvature, and matter functions in the variables $\tilde{r}=r/R_S$, $\tilde{R}=R/R_S$.
\subsection{Metric functions and Ricci scalar}
As noted in the introduction, the metric of our dynamical solution is formally the same as the metric for the static solution at radius $R$. The metric function $g_{t t}(t,r)$ is
\begin{align}
    g_{t t}(t,r)=\begin{cases}-\frac{1}{4}(3 a-b)^2=-\frac{1}{4}\left(3\sqrt{1-\frac{1}{\tilde{R}}}-\sqrt{1-\frac{\tilde{ r}^2}{\tilde{R}^3}}\right)^2,\qquad &r<R,\\
     -\left(1-\frac{1}{\tilde{r}}\right), & r\ge R.
    \end{cases}
\end{align}
 The metric function $g_{t t}$ is negative everywhere except it goes to 0 when $3a=b$, which happens on the infinite pressure surface $\tilde{r}=\tilde{R}\sqrt{9-8\tilde{R}}$. The interior $g_{t t}$ connects continuously with continuous derivatives to the Schwarzschild exterior $g_{t t}$ except at  $\tilde{r}=\tilde{R}=1$.

The metric function $g_{r r}$ is 
\begin{align}
    g_{r r}(t,r)=\begin{cases} b^{-2}=\frac{1}{1-\frac{ \tilde{r}^2}{\tilde{R}^3}}, \qquad &r<R,\\
    \left(1-\frac{1}{\tilde{r}}\right)^{-1}, &r\ge R.
    \end{cases}
\end{align}
It is always positive and goes to infinity at $\tilde{r}=\tilde{R}=1$. It does not have any special behavior connected to the infinite pressure surface. The interior $g_{r r}$ connects continuously to the Schwarzschild exterior $g_{r r}$ but $g_{r r}$ does not have continuous derivatives.

The function $\sqrt{-g(t,r)}$ is
\begin{equation}
    \sqrt{-g(t,r)}=\sqrt{-g_{t t}g_{r r}r^4 \sin^2\theta}.
\end{equation}
It goes to zero on the infinite pressure surface indicating a coordinate singularity.

\begin{figure}
    \centering
    \includegraphics[width=8cm]{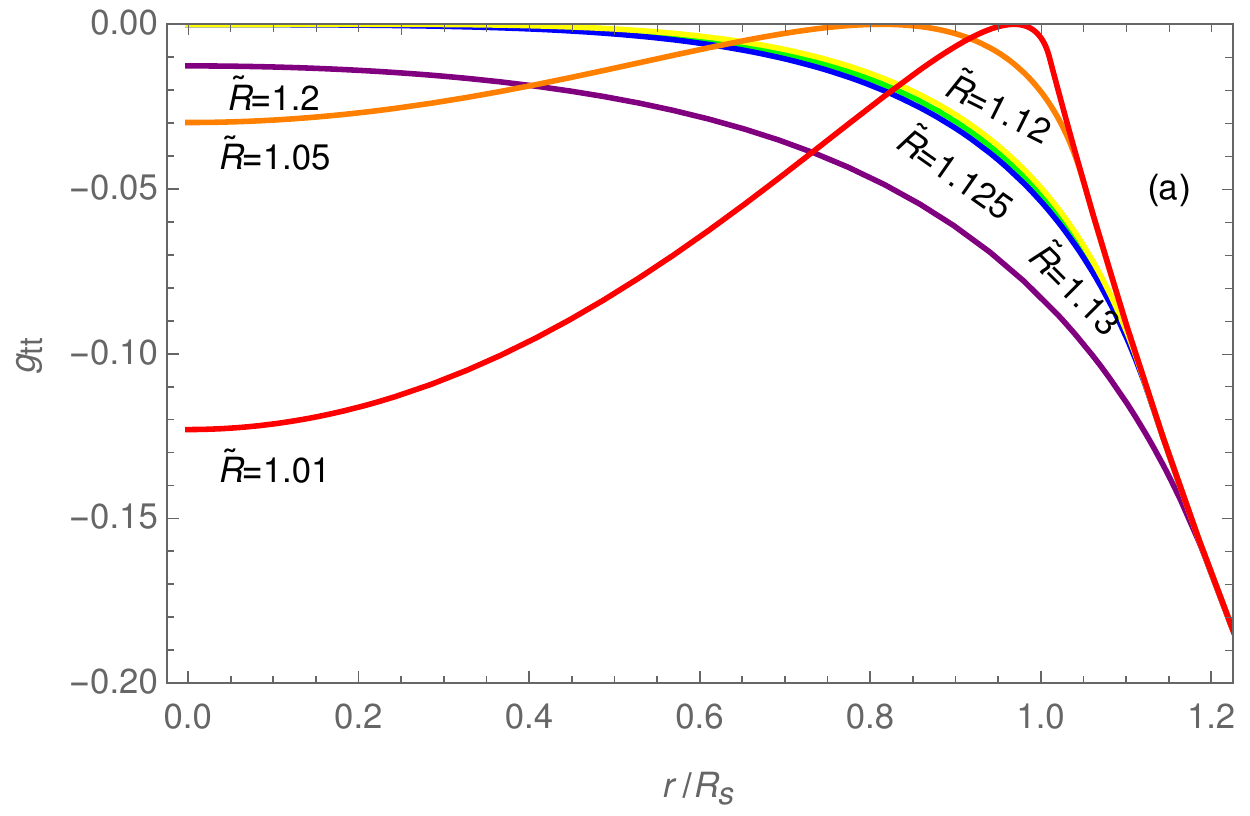}
    \includegraphics[width=8cm]{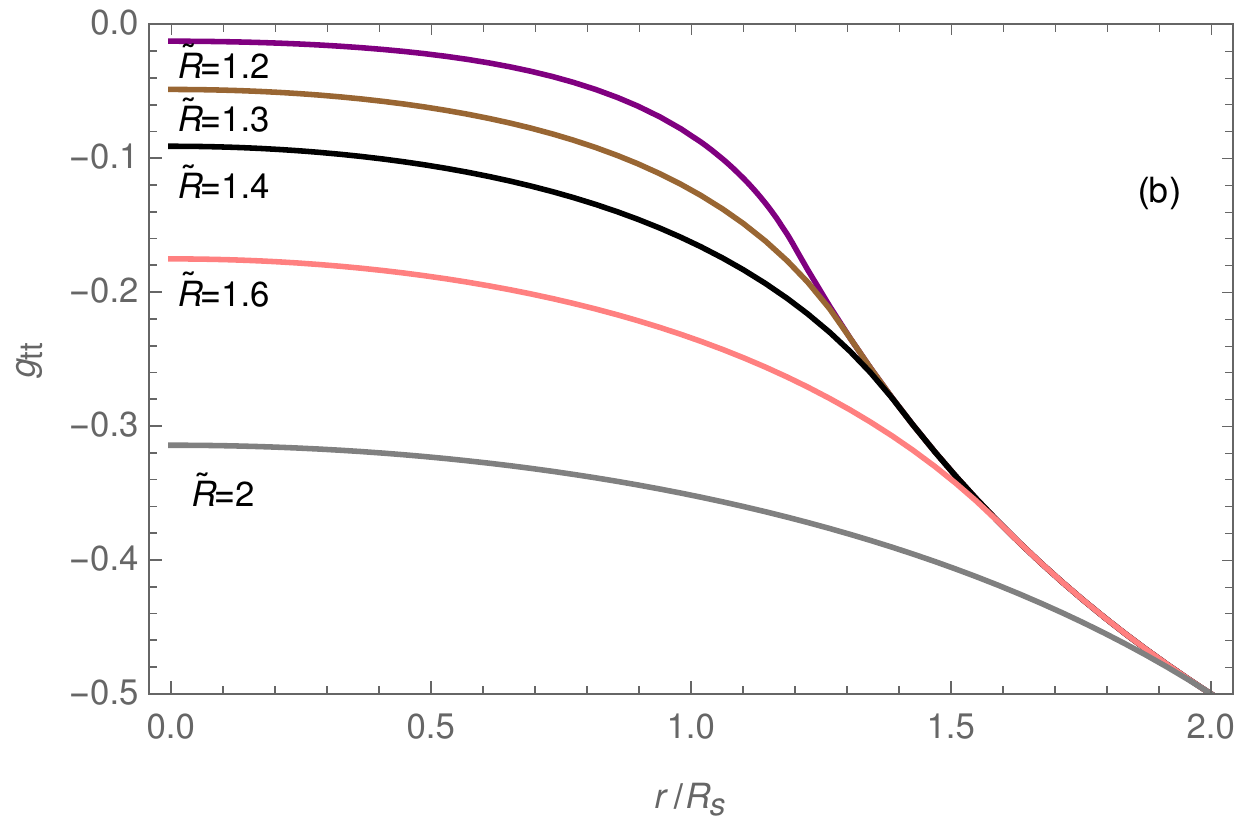}
     \includegraphics[width=8cm]{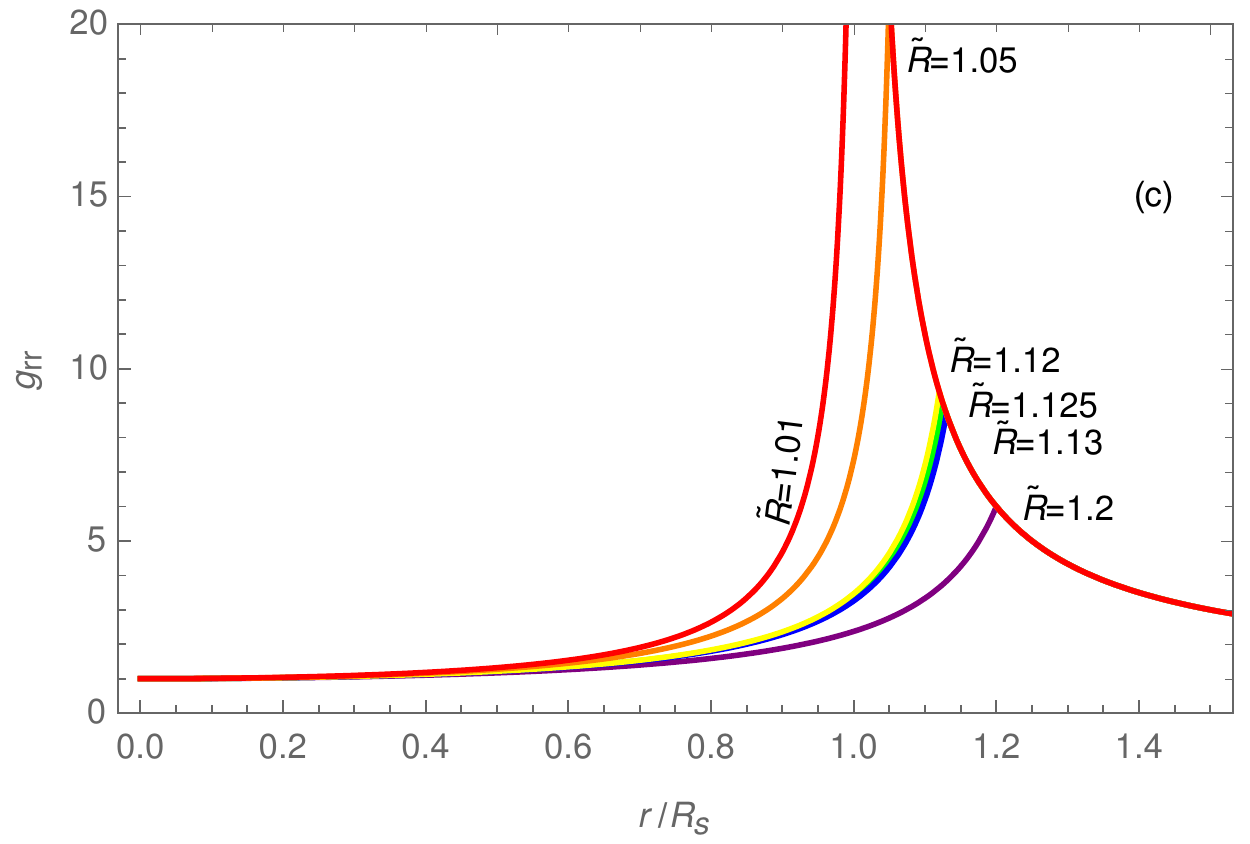}
      \includegraphics[width=8cm]{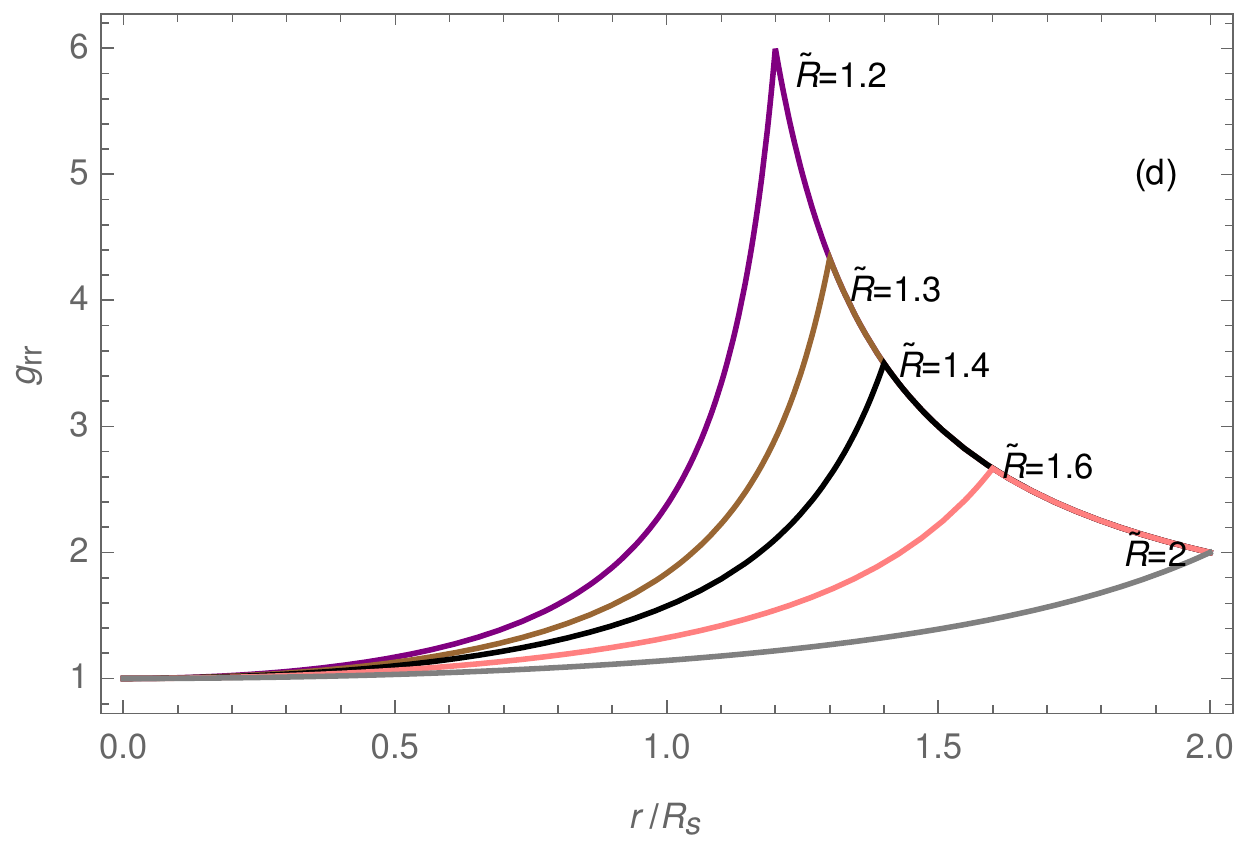}
       \includegraphics[width=8cm]{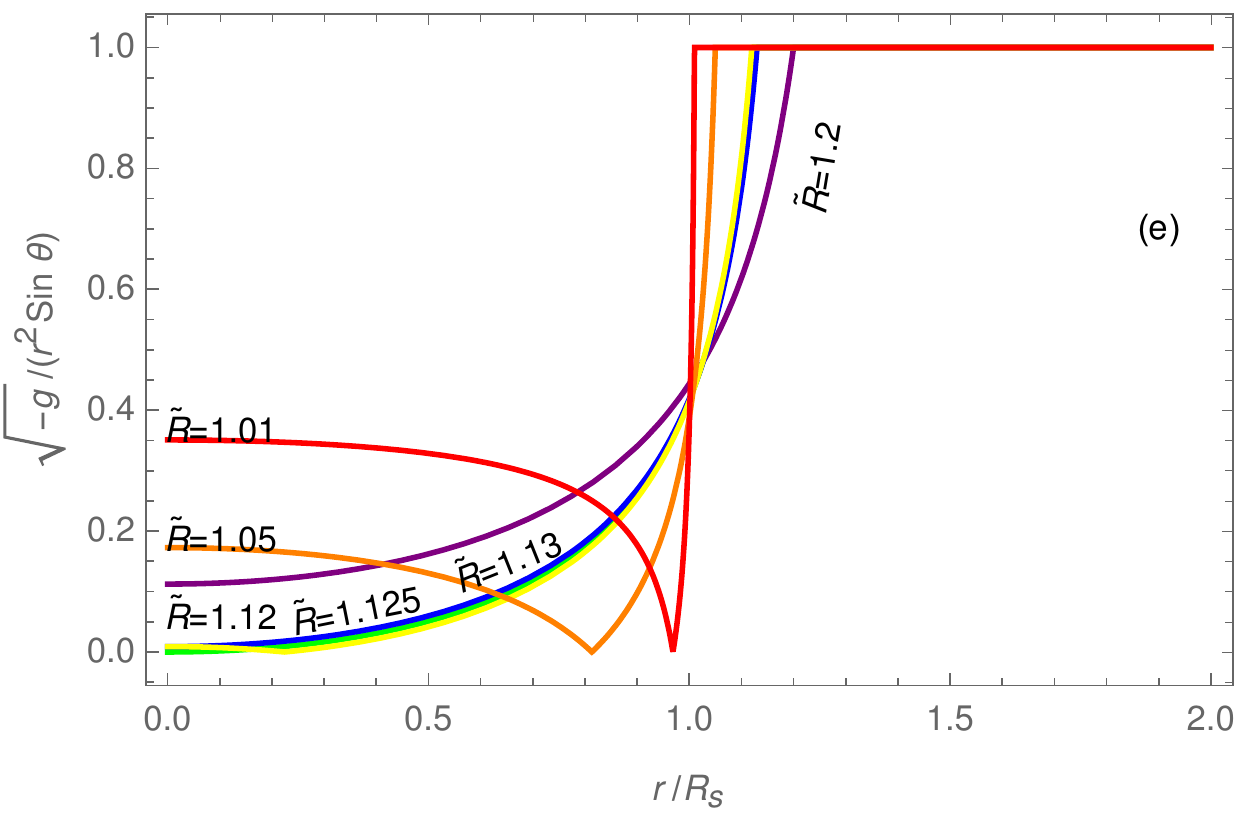}
      \includegraphics[width=8cm]{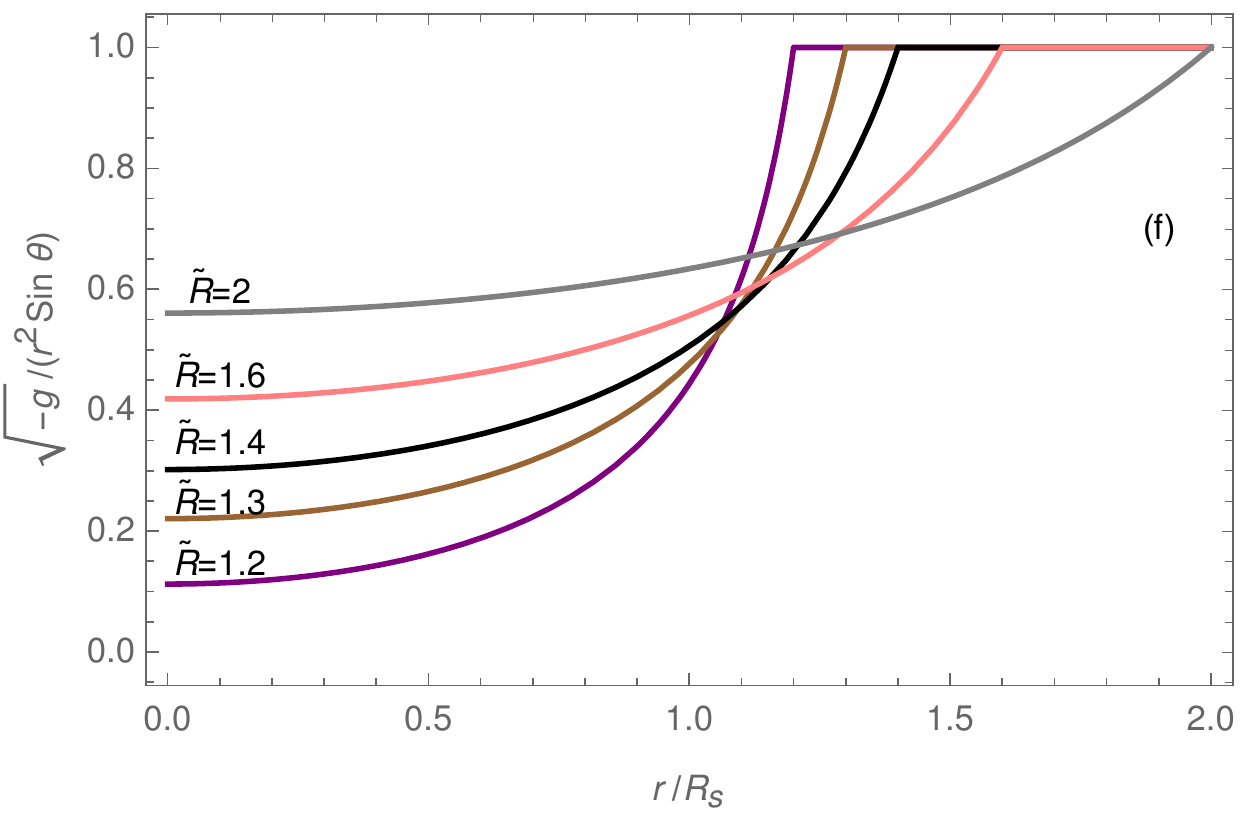}
    \caption{Plots of the metric functions $g_{t t}$ (panels (a) and (b)), $g_{r r}$ (panels (c) and (d)), and $\sqrt{-g}$ (panels (e) and (f), normalized over the $r^2 \sin\theta$ spherical factor) for various star radii $R$. Panels (a), (c), and (e) depict $R\leq 1.2 R_S$; panels (b), (d) and (f) depict $R\geq 1.2 R_S$. Profiles are labeled by $\tilde{R}=R/R_S$. These are the same as for the static solution of the same radius. Note that when $\tilde{R}\le9/8$, $g_{t t}$ and $\sqrt{-g}$ go to 0 at the radius of the infinite pressure surface. Also, as $\tilde{R}\rightarrow 1$, the value of $g_{r r}$ at $\tilde{r}=1$ goes to infinity.}
    \label{metricfuncs}
\end{figure}

\begin{figure}[H]
    \centering
    \includegraphics{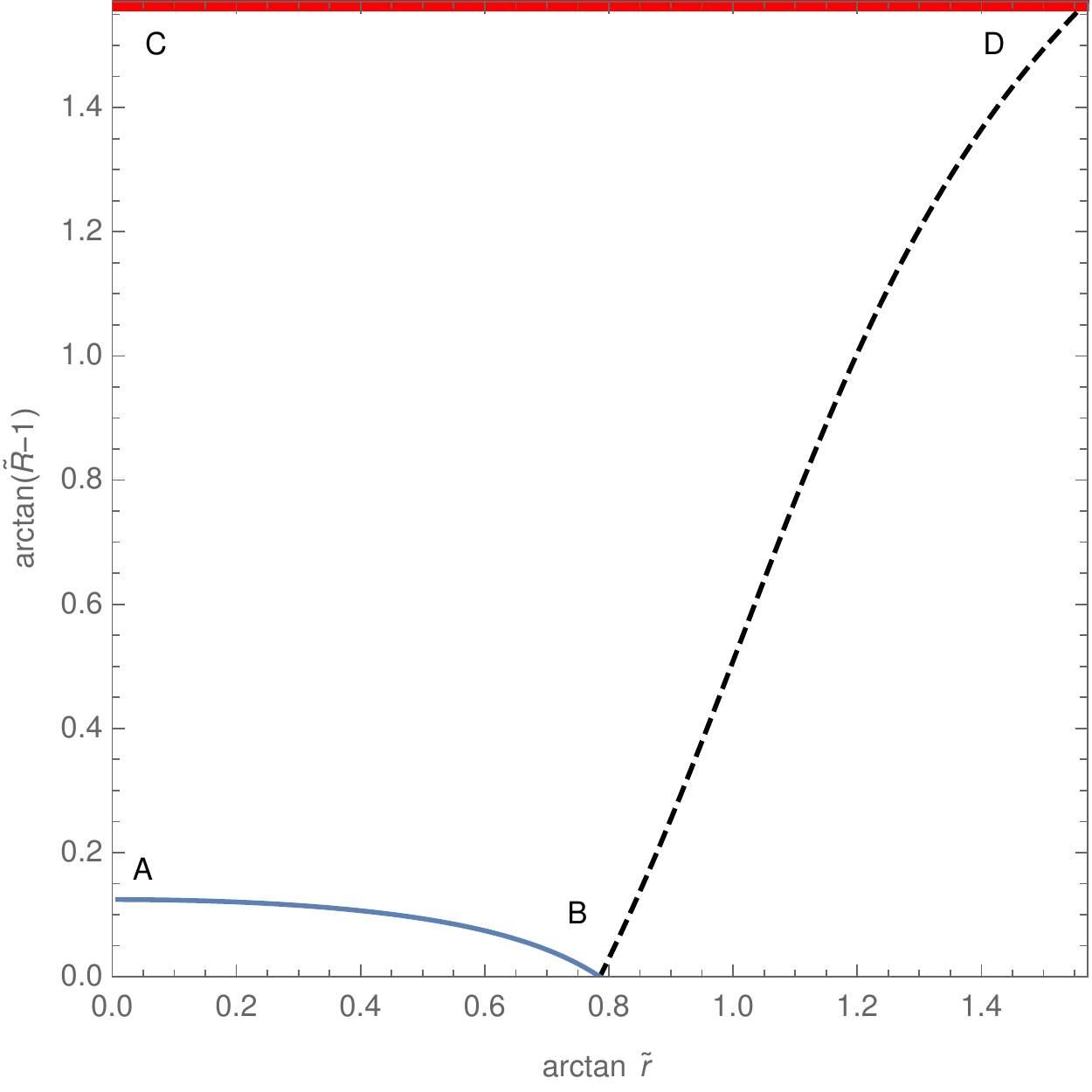}
    \caption{ Diagram showing the singularities of the Ricci scalar and the pressure anisotropy $\Delta$. The singularity associated with the infinite pressure surface is the blue line bounded by A and B. The singularity at $R\rightarrow\infty$ is the red line joining C and D. The dotted black line is the surface of the object on which the anisotropy is set to 0 by our boundary condition.}
    \label{spdiagram}
\end{figure}

Profiles of the metric functions $g_{t t}$, $g_{r r}$, and $\sqrt{-g}$ are depicted in Fig.~\ref{metricfuncs}.
The Ricci scalar is given by the expression 
\begin{equation}
    R=-8\pi G(\rho+3 p_r+2 \Delta).
\end{equation} 
It diverges on the infinite pressure surface, due to divergences in $p_r$ and $\Delta$, and at $R\rightarrow \infty$ ($t=t_0$) due to $\Delta$. These are therefore curvature singularities. The locations of the singularities in coordinates $(\arctan \tilde{r},\arctan(\tilde{R}-1))$ is depicted in Fig.~\ref{spdiagram}. The first line of singularities follows the infinite pressure surface. Its endpoints are $A=(0, \arctan \frac{1}{8})$, corresponding to $\tilde{r}=0$, $\tilde{R}=9/8$, and $B=(\pi /4, 0)$, corresponding to $\tilde{r}=1$, $\tilde{R}=1$. The second line of singularities is at $R\rightarrow\infty$. Its endpoints are $C=(0,\pi/2)$ and $D=(\pi/2, \pi/2)$.

\subsection{Energy density and radial pressure}
\begin{figure}[H]
    \centering
    \includegraphics[width=8cm]{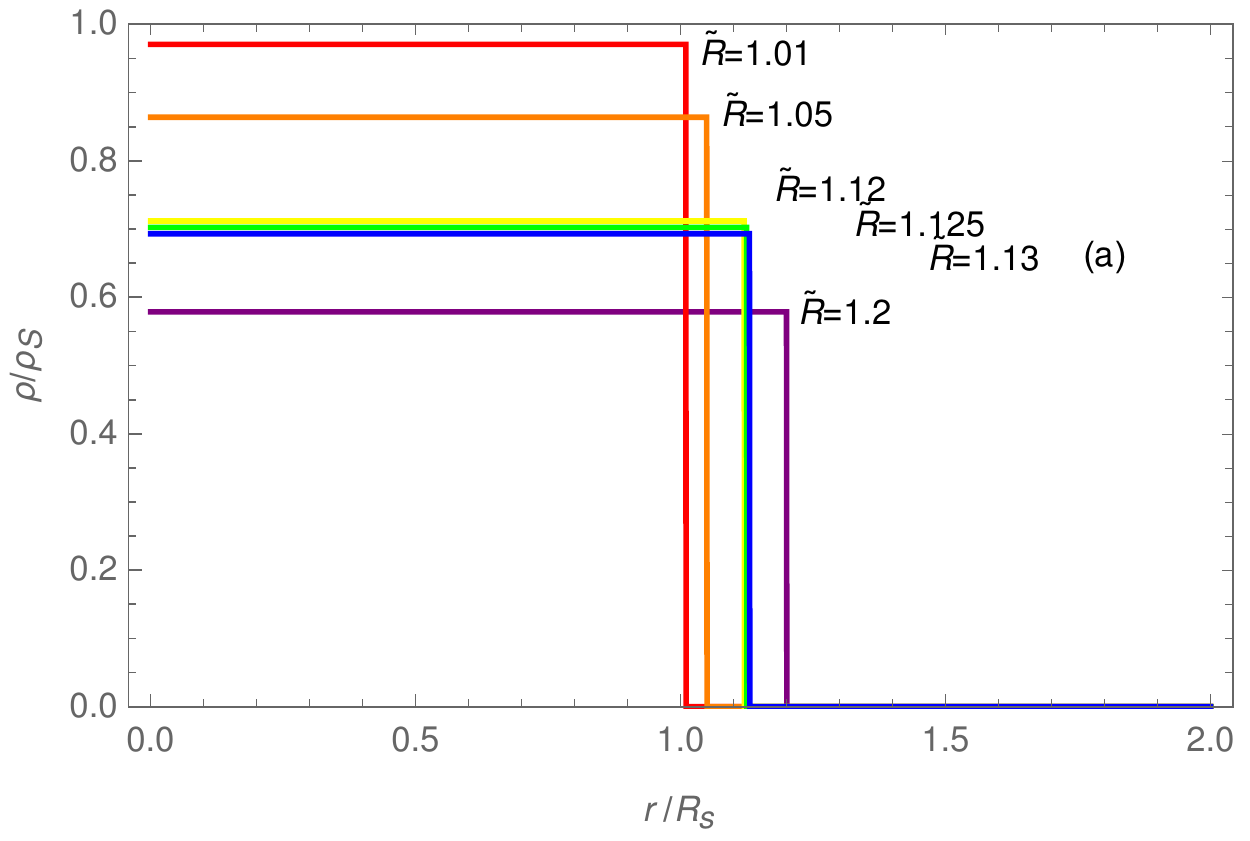}
    \includegraphics[width=8cm]{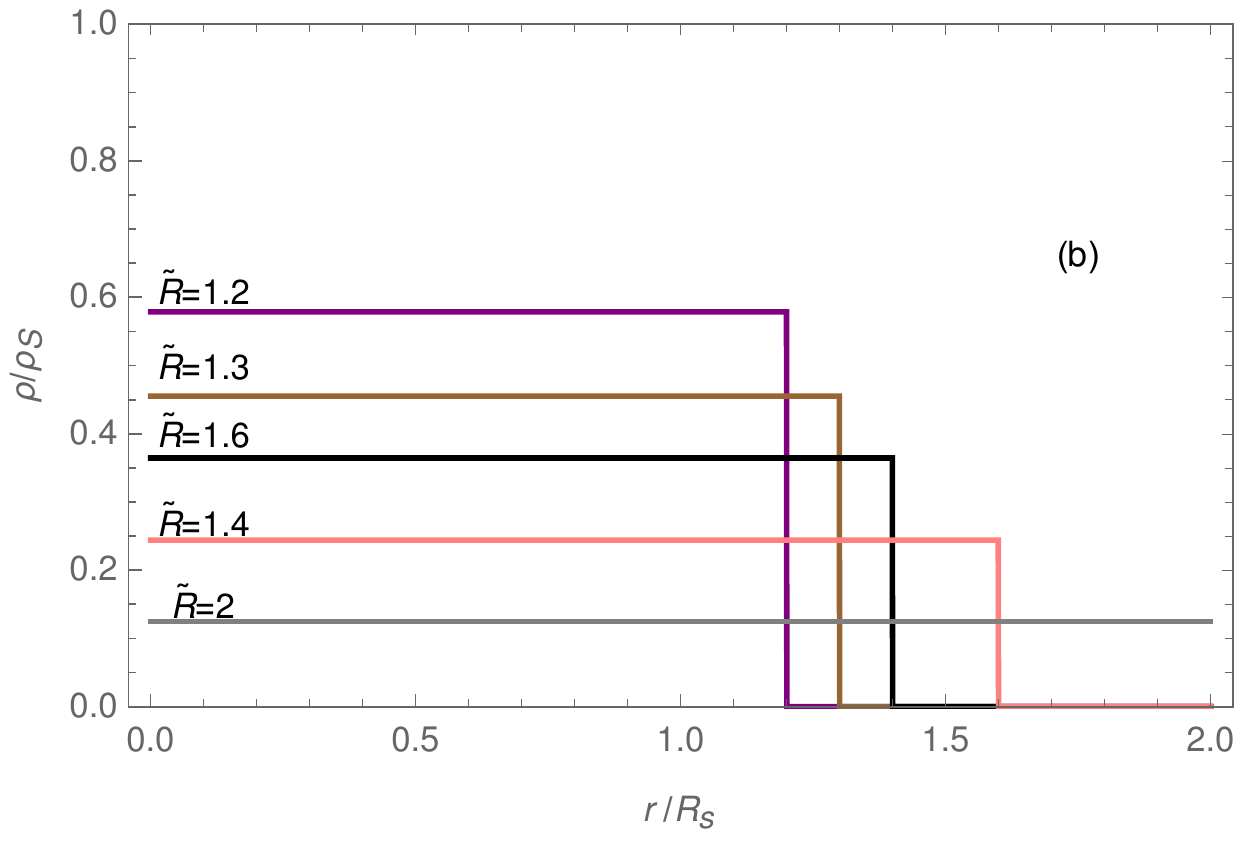}
     \includegraphics[width=8cm]{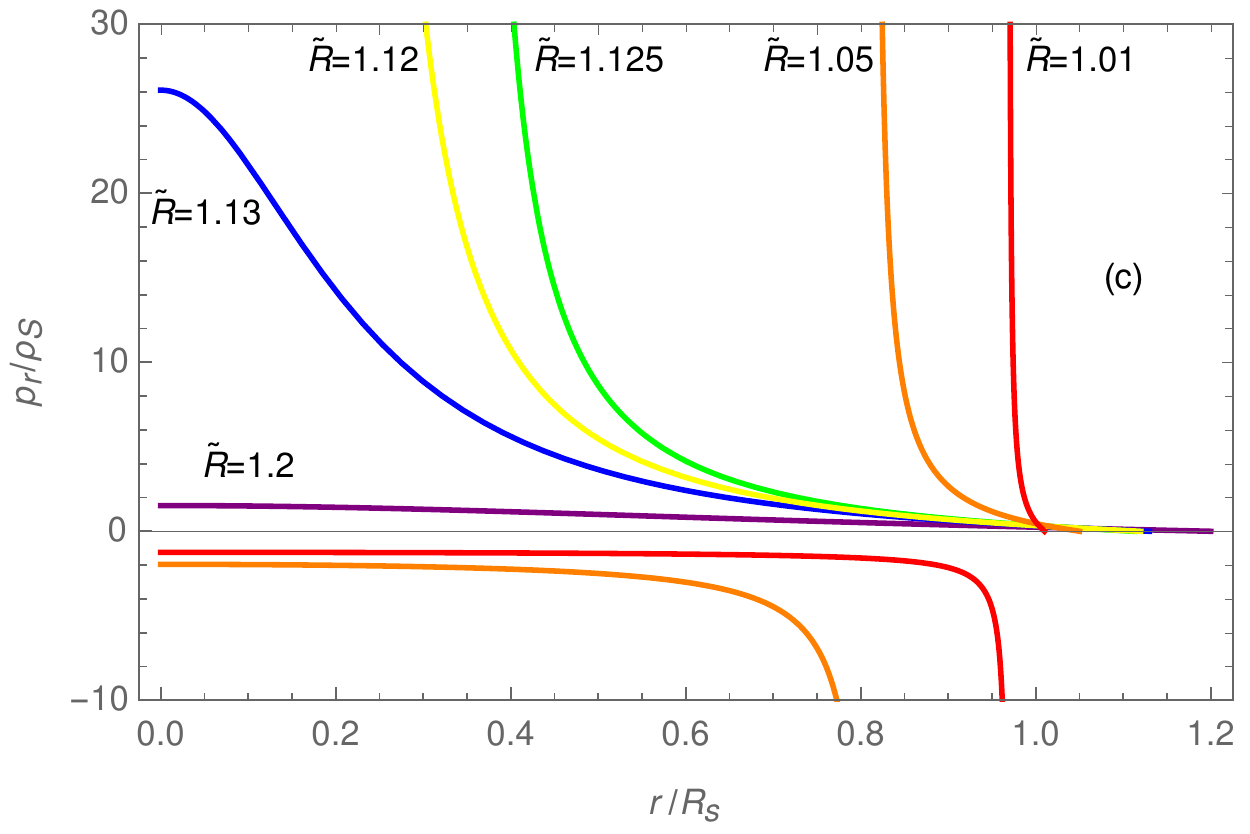}
      \includegraphics[width=8cm]{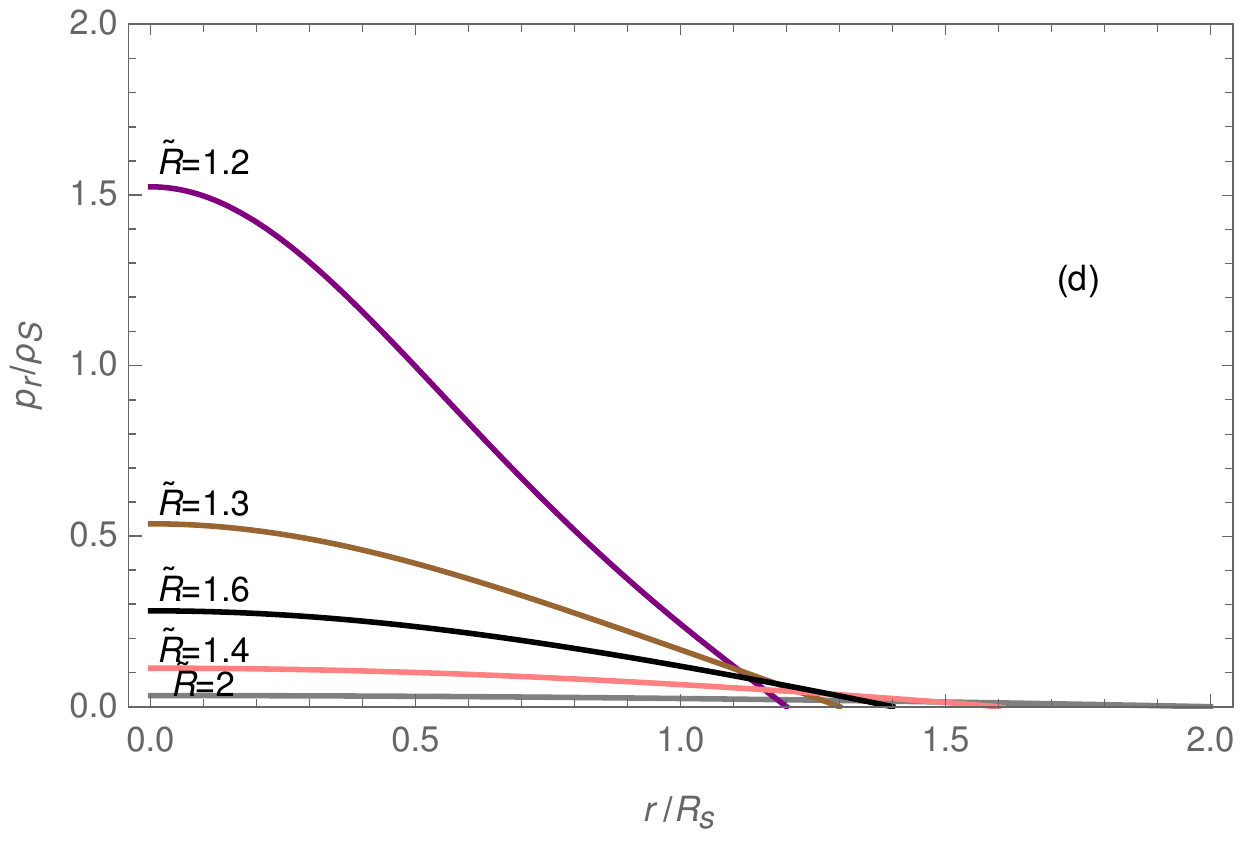}
    \caption{Density (panels (a) and (b)) and radial pressure (panels (c) and (d)) functions for various star radii $R$. Panels (a) and (c) depict $R\leq 1.2 R_S$; panels (b) and (d) depict $R\geq 1.2 R_S$. Profiles are labeled by $\tilde{R}=R/R_S$. These are the same as for the static solution of the same radius. Note the surface of infinite pressure present when $R\le9/8 R_S$ and the negative pressure region inside of it.}
    \label{sameasstat}
\end{figure}

The radial pressure $p_r(t,r)$ and energy density $\rho(t,r)$ profiles for our dynamical solution are the same as in the static solution at any $R$. For completeness we show them in Fig~\ref{sameasstat}. The density profile is constant inside the object and zero outside. The radial pressure $p_r$ is small compared to the density $\rho$ at large radii $R$. At smaller $R$, $p_r$ at the center becomes larger than $\rho$, eventually diverging when $\tilde{R}=9/8$. For $9/8>\tilde{R}>1$, the pressure is negative ($p_r<-\rho$) at the center and has a divergence at the surface of infinite pressure. This $p_r<-\rho$ behavior causes violation of the weak and null energy conditions. In the $\tilde{R}\rightarrow 1$ limit, the pressure everywhere in the interior approaches a constant value $-\rho_s$, and the surface of infinite pressure moves to the surface of the star.

\subsection{Pressure anisotropy}
\begin{figure}[H]
    \centering
   \includegraphics[width=8cm]{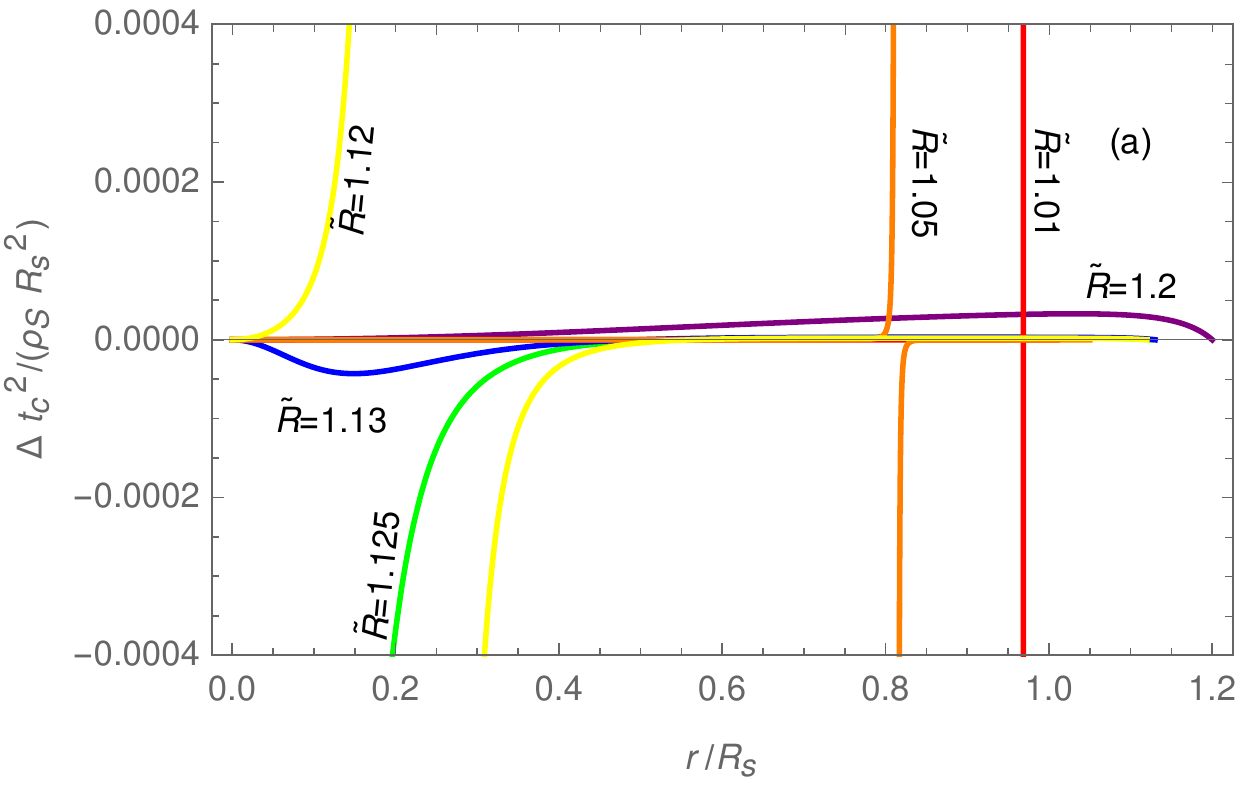}
   \includegraphics[width=8cm]{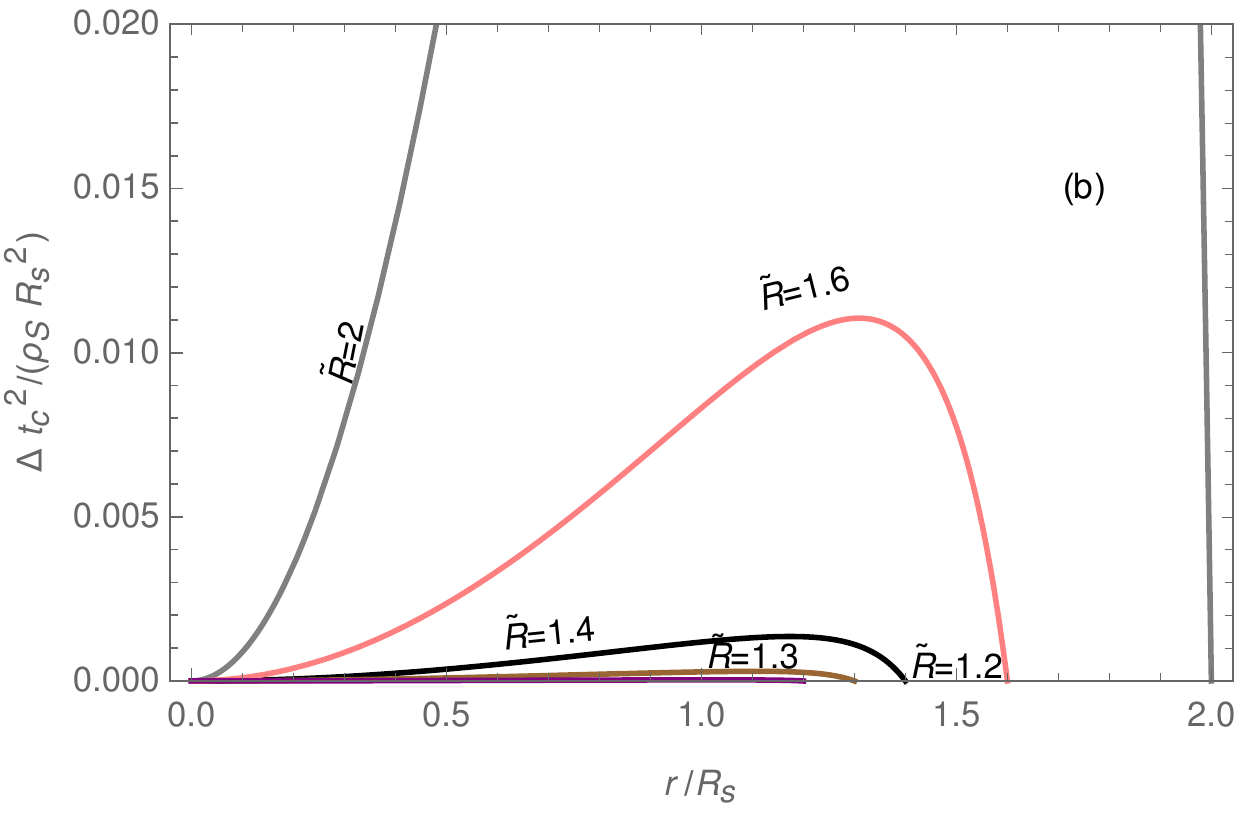}
    \caption{Rescaled anisotropy function at the same stellar radii as in Fig.~\ref{sameasstat}. Panel (a) depicts $R\leq 1.2 R_S$; panel (b) depicts $R\geq 1.2 R_S$. Profiles are labeled by $\tilde{R}=R/R_S$. For larger $R$ the anisotropy becomes roughly a quartic polynomial and rapidly decreases with decreasing $R$. For $R$ near $R_S$, the anisotropy is heavily localized near the pole in radial pressure.}
    \label{deltas}
\end{figure}

A convenient dimensionless quantity related to the pressure anisotropy function $\Delta(t,r)$ is $\tilde{\Delta}=\Delta t_c^2/(\rho_s R_S^2)$. We plot it in Fig.~\ref{deltas}. 
 The function $\Delta$ has two lines of singularities, which are lines AB and CD shown in Fig.~\ref{spdiagram}. 
The function $\tilde{\Delta}$ assumes the following limiting form near the infinite pressure surface (line AB)
\begin{equation}
    \tilde{\Delta}\approx -\frac{6\tilde{R}_0(\tilde{R}_0-1)^9(4\tilde{R}_0-3)(9-8\tilde{R}_0)}{[3(4\tilde{R}_0-1)\delta \tilde{R}+ \sqrt{9-8\tilde{R}_0}\delta \tilde{r}]^3},\qquad\text{near line AB}, \label{deltanearab}
\end{equation}
where $\delta \tilde{R}=\tilde{R}-\tilde{R}_0$, $\delta \tilde{r}=\tilde{r}-\tilde{R}_0\sqrt{9-8\tilde{R}_0}$. Thus the line AB is a line of third order poles $1/y^3$ in $\tilde{\Delta}$ except at its endpoints, where the numerator of Eq.~(\ref{deltanearab}) goes to 0 and $\tilde{\Delta}$ has essential singularities. Near point A
\begin{align}
    \tilde{\Delta}\approx\frac{\tilde{r}^2}{2^{12} 3^6(9-8\tilde{R})^3},\qquad\text{ near }\tilde{r}=0,\tilde{R}=\frac{9}{8}. \label{delA}
\end{align}
This is an essential singularity of type $x^2/y^3$. Near point B,
\begin{align}
     \tilde{\Delta}\approx\frac{6(\tilde{R}-1)^9}{(4-\tilde{r}-3\tilde{R})^3} +\frac{8(\tilde{R}-1)^8}{(4-\tilde{r}-3\tilde{R})^2}-\frac{2(\tilde{R}-1)^7}{3(4-\tilde{r}-3\tilde{R})}+(\tilde{R}-1)^7 g(\tilde{r},\tilde{R}),\text{ near }\tilde{r}=1,\tilde{R}=1, \label{delB}
\end{align}
where $g(\tilde{r},\tilde{R})$ is a nonsingular function well approximated by
\begin{align}
   & g(\tilde{r},\tilde{R})\approx \frac{2}{3(C \tilde{R}-\tilde{r}-1-C)}-\frac{A(\tilde{R}-1)^3}{(B \tilde{R}-\tilde{r}+1-B)^4},\\
    & B=1.8629316,\qquad C=3.5717930,\qquad A=\frac{(9+55C)(B-1)^4}{96(C-1)}.
\end{align}
The expression for $A$ ensures $\tilde{\Delta}=0$ at $r=R$. Here there are essential singularities of the type $x^9/y^3$, $x^8/y^2$, and $x^7/y$.

Near the line CD ($R\rightarrow \infty$) the limiting form of $\tilde{\Delta}$ is
 \begin{equation}
       \tilde{\Delta} \approx \frac{15}{8} \tilde{r}^2(\tilde{R}^2-\tilde{r}^2),  \qquad  \tilde{R}\gg1.
   \label{bigbigD}
 \end{equation}
 This diverges with $\tilde{R}^2$ except at points C and D which again are essential singularities.
\subsection{Momentum Density}
\begin{figure}[H]
    \centering
   \includegraphics[width=8cm]{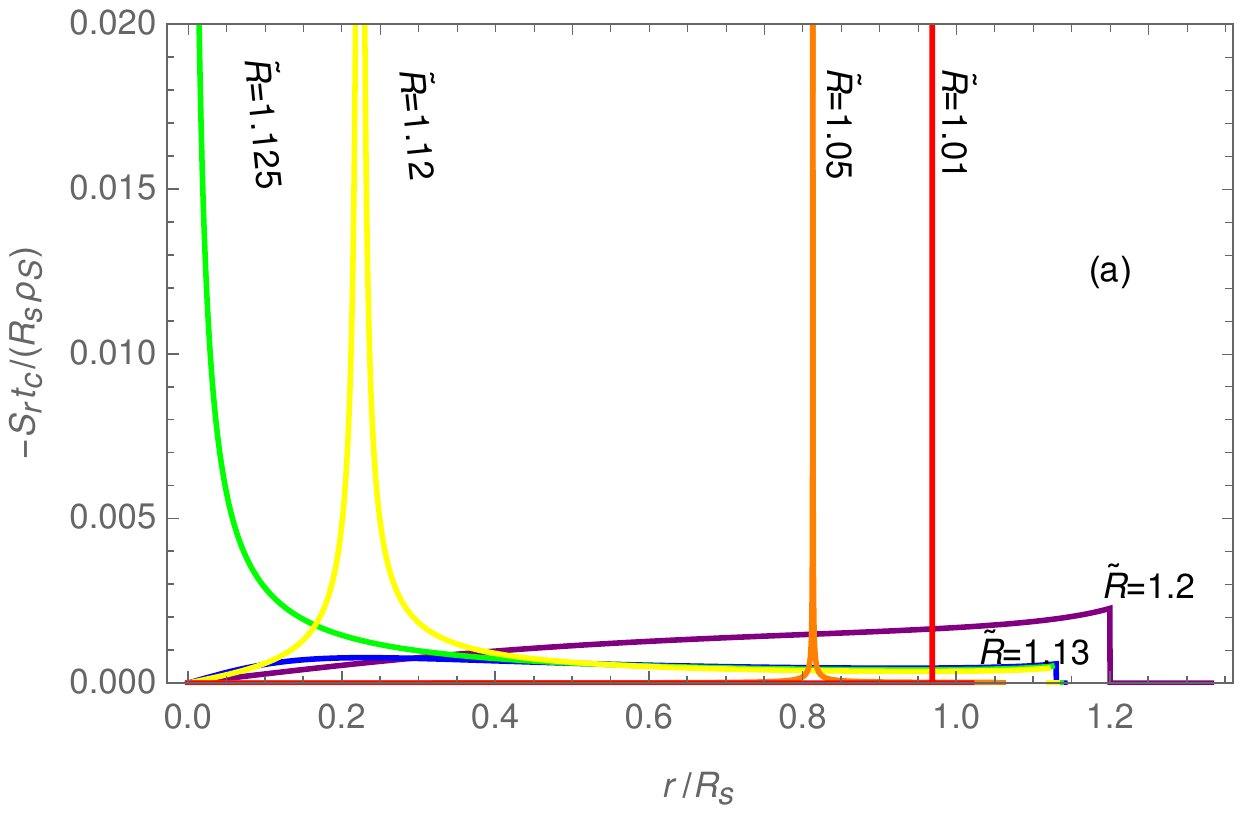}
   \includegraphics[width=8cm]{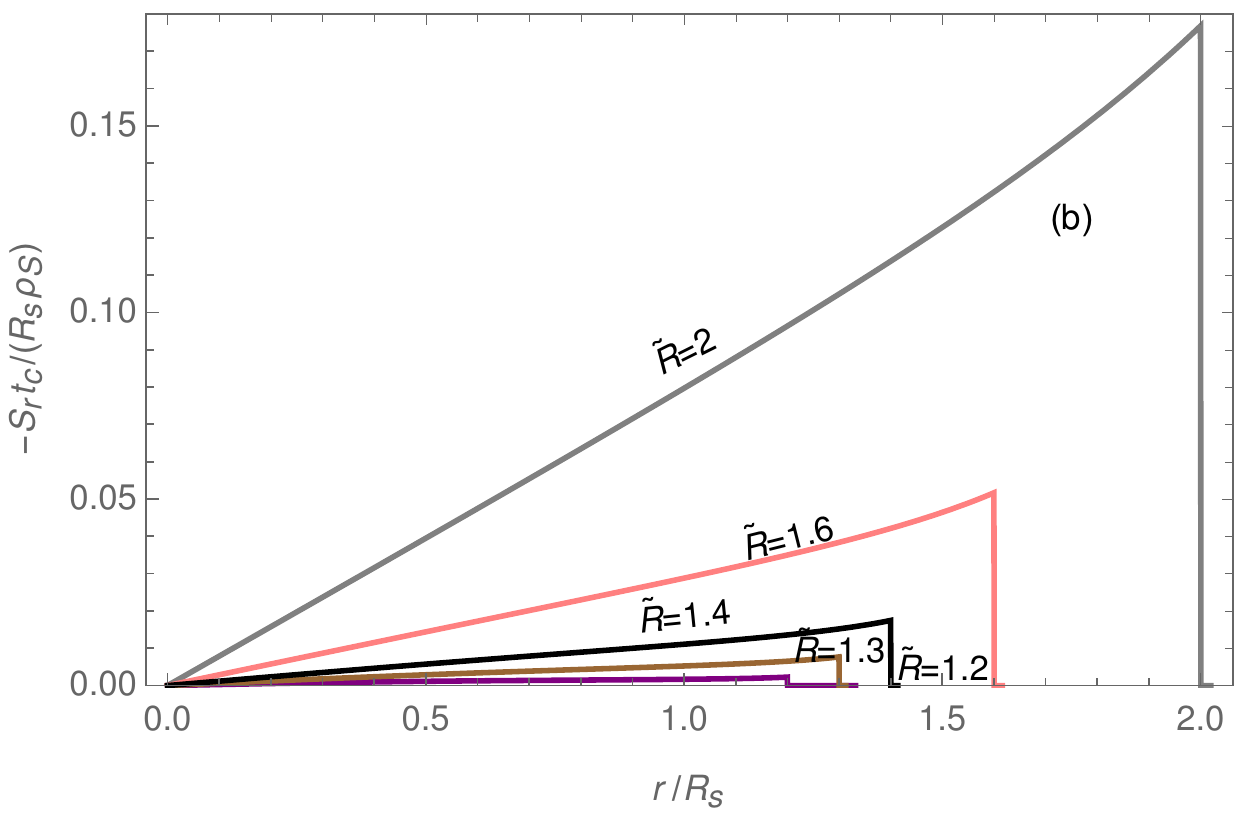}
    \caption{Rescaled momentum function at various star radii $R$.  Panel (a) depicts $R\leq 1.2 R_S$; panel (b) depicts $R\geq 1.2 R_S$. Profiles are labeled by $\tilde{R}=R/R_S$. For $R$ near $R_S$, the momentum is heavily localized near the infinite pressure surface.}
    \label{srs}
\end{figure}

A convenient dimensionless quantity related to $S_r$ is $\tilde{S_r}=S_r t_c/(\rho_s R_S)$. Figure \ref{srs} shows plots of $-\tilde{S_r}$.

The function $S_r$ can also be examined at the points on Fig.~\ref{spdiagram}. The limiting forms are

\begin{align}
    \tilde{S_r}\approx -\frac{2\sqrt{9-8\tilde{R}_0}(\tilde{R}_0-1)^{9/2}}{\tilde{R}_0^{3/2}|\sqrt{9-8\tilde{R}_0}\delta \tilde{r}+3(4\tilde{R}_0-3)\delta \tilde{R}|}\qquad\text{(near line AB)}\label{SrABline}
\end{align}
 again using $\delta \tilde{R}=\tilde{R}-\tilde{R}_0$, $\delta \tilde{r}=\tilde{r}-\tilde{R}_0\sqrt{9-8\tilde{R}_0}$.
\begin{align}
    &\tilde{S_r}\approx\frac{-\tilde{r}}{2^1 3^7|8\tilde{R}-9|} \qquad &&\tilde{R}=9/8,\tilde{r}=0\text{ (near point A)};\label{SrA}\\
    &\tilde{S_r}\approx\frac{2(\tilde{R}-1)^{9/2}}{|4-3\tilde{R}-\tilde{r}|}+h(\tilde{r},\tilde{R}),\qquad &&\tilde{R}=1,\tilde{r}=1\text{ (near point B)};\label{Srb}\\
    &\tilde{S_r} \approx-\tilde{r},\qquad  &&\tilde{R}\gg1\text{ (near line CD)}.
   \label{bigbigSr}
\end{align}
Here $h(\tilde{r},\tilde{R})$ is a nonsingular function. We see from Eq.~(\ref{SrABline}) that the line AB is a line of singularities of type $1/|y|$ in $S_r$. From Eq.~(\ref{SrA}), $S_r$ has a singularity of type $x/|y|$ at point A, and from Eq.~(\ref{Srb}), $S_r$ has a singularity of type $x^{9/2}/|y|$ at point B. The line CD is not singular except at point D where $\tilde{S_r}$ diverges as $-\tilde{R}$.
\subsection{Force analysis}
In  \cite{Beltracchi:2018ait} we found that general time dependent spherically symmetric systems satisfy a force equation
\begin{equation}
    - \frac{\partial p_r}{\partial r}-\frac{G \left( m+4\pi r^3 p_r \right) \left( \rho+p_r \right)}{r^2 \left( 1 - \frac{2 G m}{r} \right) }+\frac{2\Delta}{r}=\sqrt{1-\frac{2Gm}{r}}   \frac{1}{\sqrt{-g_{t t}}}\frac{\partial}{\partial t} \left( \frac{S_r}{1-\frac{2Gm}{r}} \right).
     \label{feq}
\end{equation}
This is a dynamical anisotropic generalization of the Tolman--Oppenheimer--Volkoff equation. The right hand side of Eq.~(\ref{feq}) must be 0 for static systems. Here $m$ is the mass inside radius $r$. For our solution, 
\begin{align}
    m=\begin{cases}
    M\frac{r^3}{R^3},\qquad &r<R,\\
    M,\qquad &r\ge R.
    \end{cases}
\end{align}
The static Schwarzschild interior solution satisfies the standard isotropic Tolman--Oppenheimer--Volkoff equation at all points with finite pressure\footnote{It is argued in  \cite{0264-9381-32-21-215024} that the $dp_r/dr$ term produces a Dirac delta function at the infinite pressure surface, which is compensated for by another Dirac delta function in the anisotropy term. }.
This means that the only terms that survive in the time-dependent force equation are the anisotropy force and the changing momentum terms. Hence Eq.~(\ref{feq}) reduces to Eq.~(\ref{eq:ptA}).

For the collapsing solution ($t_c>0$) the anisotropy force acts as a force to slow down the initially rapid collapse. As the center pressure is diverging, the anisotropy force pulls inward, see the $\tilde{R}=1.13$ and $\tilde{R}=1.125$ curves in Fig.~\ref{deltas}. At late times, the anisotropy is positive inside the pressure divergence and negative outside; this indicates that the anisotropy force is pulling into the divergence rather than pushing away.

For the expanding solution ($t_c<0$), the anisotropy force is the same at the same values of $R$, but the momentum term $S_r$ has opposite sign and increases, rather than decreases, and the object expands rather than contracts.  
\subsection{Energy-momentum tensor type}
As mentioned in section II, the energy-momentum tensor $T_{\mu \nu}$ is type I when $(\rho+p_r)^2-4 S_r^2\ge0$ and is type IV otherwise. Using the expressions from Eqs.~(\ref{rhoandpr}) and (\ref{eq:Sr}) we can obtain a condition for where in the $(r,R)$ plane $T_{\mu \nu}$ is type I and where it is type IV. It is type I when
\begin{equation}
    r\le r_{IV}=\frac{t_c\tilde{R}^{3/2}}{\sqrt{t_c^2/R_S^2+(\tilde{R}-1)^8\tilde{R}}},
\end{equation}
and type IV otherwise. Energy-momentum tensors of type IV cannot satisfy the weak energy condition \cite{Hawking:1973uf}. Note that $T_{\mu \nu}$ is type I at $r=0$ for all times, but the outer region of the object where there is more momentum is type IV if $r_{IV}\le r <R$. The type IV outer region shrinks or grows with $R$ and disappears when $R\le r_{IV}$, i.e.,
\begin{equation}
    4\tilde{R}(1-\tilde{R})^7\le\frac{t_c^2}{R_S^2}.
\end{equation}
\section{Conclusion}
The interior Schwarzschild solution, despite its perhaps unnatural uniform density, still has interesting properties such as the Buchdahl limit, the gravastar limit, and the extended Kerr source. In this paper, we generalize the interior Schwarzschild solution to include collapse or expansion. Our solution to the Einstein field equations is interesting mathematically since it is exact and fairly simple, allowing for a detailed analysis of its features. Our expanding solution starts as a sphere of dark energy in the infinite past and reaches an infinite size at $t_0$. Our collapsing system starts at an infinite size and asymptotically approaches a sphere of dark energy at large times. Therefore our collapsing solution may be thought of as a kind of formation process for gravastars or dark energy stars, which in the terminology of \cite{Beltracchi:2018ait} are astrophysical objects with a dark energy core. However, our collapsing solution involves spacetime singularities and violations of the weak and null energy conditions. Other formation processes that are nonsingular and do not violate those energy conditions exist \cite{Beltracchi:2018ait}. It would be interesting to see if other static or stationary dark energy stars \cite{Bardeen,Dymnikova1992,Dymnikova2003,Dymnikova:2000zi,Dymnikova:2001mb,DYMNIKOVA2007358,Dymnikova:2001fb,Dymnikova:2015yma,mazur2001gravitational,MM2004,Visser:2003ge,cattoen2005gravastars,doi:10.1080/13642810108221981,Chapline:2005ph} can be generalized to include a nontrivial time dependence in a simple way.  

\section*{Acknowledgements}
 P.G. thanks Emil Mottola for an intriguing conversation that rekindled his interest in gravastars. This work has been partially supported by NSF award PHY-1720282 at the University of Utah.

\medskip

\bibliographystyle{unsrt
}
\bibliography{samples.bib}

\end{document}